%% file: main.tex
\pdfoutput=1
\UseRawInputEncoding

\documentclass[sigconf]{acmart}

\AtBeginDocument{%
  }

\copyrightyear{2025}
\acmYear{2025}
\setcopyright{acmlicensed}
\acmConference[SIGIR '25] {Proceedings of the 48th International ACM SIGIR Conference on Research and Development in Information Retrieval}{ July 13--18, 2025}{Padua, Italy.}
\acmBooktitle{Proceedings of the 48th International ACM SIGIR Conference on Research and Development in Information Retrieval (SIGIR '25), July 13--18, 2025, Padua, Italy}
\acmISBN{979-8-4007-1592-1/25/07}
\acmDOI{10.1145/3726302.3729894}

\usepackage{multirow,tabularx}
\usepackage{amsmath} 
\usepackage{enumitem}
\usepackage{algorithm}  
\usepackage{algorithmicx}
\usepackage{algpseudocode} 
\usepackage{subcaption}
\usepackage{amsfonts} 
\usepackage{bm}
\PassOptionsToPackage{table}{xcolor}
\usepackage[table]{xcolor}
\usepackage{colortbl}
\usepackage{booktabs}
\usepackage{graphicx}
\usepackage[most]{tcolorbox} 

\newcommand{\ie}{\emph{i.e., }}
\newcommand{\eg}{\emph{e.g., }}

\begin{document}

\title{AlphaFuse: Learn ID Embeddings for Sequential Recommendation in Null Space of Language Embeddings}


\author{Guoqing Hu}
\email{HugoChinn@mail.ustc.edu.cn}
\orcid{0009-0004-4089-7606}
\affiliation{
\institution{University of Science and Technology of China}
\city{Hefei}
\country{China}
}

\author{An Zhang}
\email{an_zhang@nus.edu.sg}
\orcid{0000-0003-1367-711X}
\affiliation{
  \institution{National University of Singapore}
  \department{School of Computing}
  \country{Singapore}}

\author{Shuo Liu}
\email{shuoliu@stu.ecnu.edu.cn}
\orcid{0000-0001-7970-3187}
\affiliation{%
  \institution{East China Normal University}
  \city{Shanghai}
  \country{China}
}

\author{Zhibo Cai}
\email{caizhibo@ruc.edu.cn}
\authornote{Corresponding author.}
\orcid{0009-0007-3582-8835}
\affiliation{%
  \institution{Renmin University of China}
  \department{Center for Applied Statistics and School of Statistics}
  \city{Beijing}
  \country{China}
}

\author{Xun Yang}
\email{xyang21@ustc.edu.cn}
\orcid{0000-0003-0201-1638}
\affiliation{%
  \institution{University of Science and Technology of China}
  \city{Hefei}
\country{China}}

\author{Xiang Wang}
\email{xiangwang1223@gmail.com}
\orcid{0000-0002-6148-6329}
\affiliation{%
  \institution{University of Science and Technology of China}
  \city{Hefei}
\country{China}}




\begin{abstract}
    Recent advancements in sequential recommendation have underscored the potential of Large Language Models (LLMs) for enhancing item embeddings.
    However, existing approaches face three key limitations:
    1) the degradation of the semantic space when high-dimensional language embeddings are mapped to lower-dimensional ID embeddings,
    2) the underutilization of language embeddings,  
    and 3) the reliance on additional trainable parameters, such as an adapter, to bridge the gap between the semantic and behavior spaces.
    In this paper, we introduce \textbf{AlphaFuse}, a simple but effective language-guided learning strategy that addresses these challenges by learning ID embeddings within the null space of language embeddings.
    Specifically, we decompose the semantic space of language embeddings via Singular Value Decomposition (SVD), distinguishing it into a semantic-rich row space and a semantic-sparse null space. 
    Collaborative signals are then injected into the null space, while preserving the rich semantics of the row space.
    AlphaFuse prevents degradation of the semantic space, integrates the retained language embeddings into the final item embeddings, and eliminates the need for auxiliary trainable modules, enabling seamless adaptation to any sequential recommendation framework.
    We validate the effectiveness and flexibility of AlphaFuse through extensive experiments on three benchmark datasets, including cold-start user and long-tail settings, showcasing significant improvements in both discriminative and diffusion-based generative sequential recommenders.
    Our codes and datasets are available at \url{https://github.com/Hugo-Chinn/AlphaFuse}.
\end{abstract}

 


\maketitle

\input{chapters/1-intro}

\input{chapters/2-prelim}
\input{chapters/3-method}

\input{chapters/4-exp}

\input{chapters/5-relawork}
\input{chapters/6-conclusion}

\bibliographystyle{style/ACM-Reference-Format}

\bibliography{ref}

\appendix

\end{document}

%% file: chapters/1-intro.tex
\section{Introduction}\label{sec:intro}
\begin{figure}[t!]
    \centering
    \includegraphics[width=0.4\textwidth]{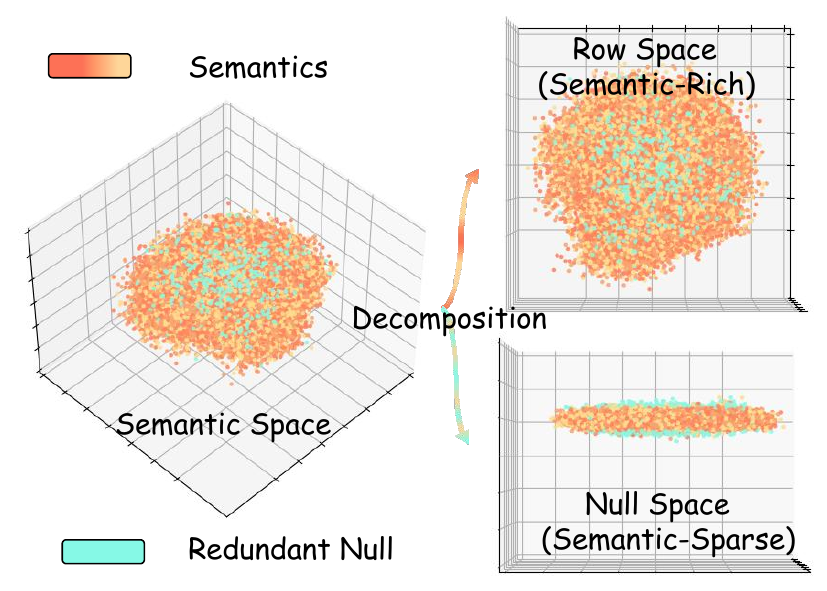}
    \vspace{-10pt}
    \caption{Visualization of space decomposition.}
    \vspace{-15pt}
    \label{fig:spaces}
\end{figure}

 \begin{figure*}[t!] 
    \centering 
    \includegraphics[width=\textwidth]{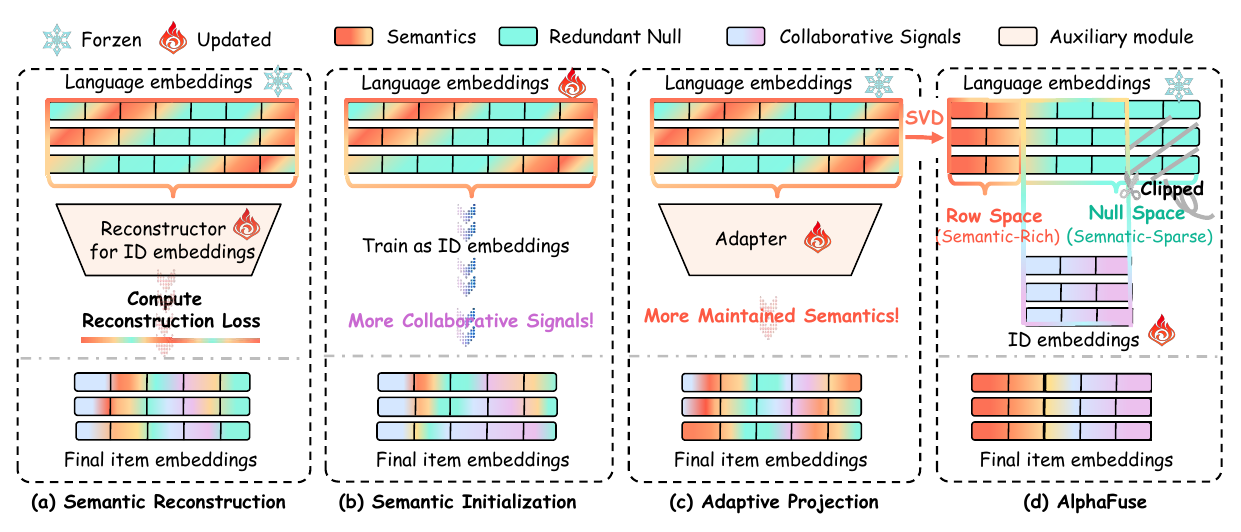}
    \vspace{-10pt}
    \caption{Overall framework of language-guided ID embedding learning strategies.}
    \vspace{-10pt}
    \label{figframework}
 \end{figure*}

Sequential recommendation focuses on predicting the next item of interest based on the sequence of items a user has interacted with.
Learning high-quality item embeddings is essential for sequential recommenders to effectively capture users' behaviors and preferences \cite{SASRec,DreamRec}.
Early works \cite{GRURec,Caser,SASRec,Bert4Rec,DreamRec} assign each item a unique ID and learn ID embeddings from scratch using user-item interactions. 
These embeddings are designed to capture collaborative signals within the interaction behavior space \cite{ID1, MoRec, WhitenRec}.
Recent works \cite{RLMRec, AlphaRec, LLMESR}, inspired by the success of LLMs \cite{GPT3,OpenAI_ada}, utilize LLMs to encode item textual metadata (\eg titles, descriptions) into language embeddings, forming a semantic space rich in world knowledge.
Embeddings derived from these two research lines encapsulate distinct types of information --- ID embeddings, typically trainable, emphasize collaborative signals among users, while language embeddings, often frozen, capture the semantic and contextual knowledge of items.
This naturally raises a question: How can language embeddings effectively guide the learning of ID embeddings and be further integrated into better item embeddings?

Several language-guided learning strategies for ID embeddings have been proposed \cite{KAR,RLMRec,MoRec,WhitenRec,AlphaRec,LLM2BERT4Rec,SAID, elephant, LLMESR}, aiming to fuse collaborative signals with semantic knowledge.
1) \textbf{Semantic reconstruction} \cite{RLMRec} initializes ID embeddings randomly and incorporates an additional module to reconstruct language embeddings, thereby enhancing representation learning in a manner similar to autoencoders \cite{AE}. 
This approach enforces behavior space to be close to semantic space.
2) \textbf{Semantic initialization} \cite{LLM2BERT4Rec,SAID,elephant} uses language embeddings to initialize ID embeddings, and then update them during training, thus deriving behavior space from semantic space.
3) \textbf{Adaptive projection} \cite{UniSRec, MoRec, WhitenRec, AlphaRec} passes language embeddings through a trainable adaptor (say, 2-layer MLP) to output ID embeddings,  thus mapping semantic space into behavior space.

Despite the effectiveness of these strategies, we scrutinize their pipelines, as Figure \ref{figframework} shows, and identify three inherent limitations:
\textbf{Degeneration of semantic space.} 
The semantic space generated by language embeddings encapsulates the rich world knowledge of LLMs and typically requires high-dimensional representations (\eg 1536 in OpenAI's text-embedding-3-small \cite{OpenAI_ada}). 
In contrast, the behavior space represented by ID embeddings is inherently lower-dimensional (\eg 64 or 128).
Consequently, deriving ID embeddings from language embeddings can cause the semantic space to degenerate into a lower-dimensional manifold, highlighting a fundamental mismatch between these spaces.
\textbf{Underutilization of language embeddings.}
Semantic reconstruction and semantic initialization only treat language embeddings as a guide for learning ID embeddings, rather than explicitly incorporating language embeddings into the final item embeddings.
Consequently, the rich world knowledge encoded in language embeddings is untouched during inference.
\textbf{Involvement of auxiliary trainable parameters.} 
Both semantic reconstruction and adaptive projection rely on auxiliary modules, such as a reconstructor or adapter, to learn ID embeddings and finalize item embeddings.
Taking an adapter as an example, a two-layer MLP mapping between the high-dimensional language embeddings and the lower-dimensional ID embeddings involves a substantial number of trainable parameters, often surpassing the complexity of the recommendation model itself.
This leads to parameter redundancy and reduced inference efficiency.

To address these limitations, we propose that ID embeddings should be semantic-anchored (\ie preserving the original semantic space of language embeddings) and tuning-efficient (\ie avoiding additional trainable parameters beyond ID embeddings). 
To achieve this, inspired by AlphaEdit \cite{AlphaEdit}, we introduce a method to learn ID embeddings within the null space of language embeddings, named \textbf{AlphaFuse}.
The pipeline consists of four steps:
(1) \textbf{Decomposition of semantic space}: 
We perform singular value decomposition (SVD) \cite{LinearA1,LinearA2} on language embeddings, dividing the semantic space into singular subspaces associated with distinct singular values. 
Subspaces with zero singular values (geometrically termed as the null space \cite{LinearA1, LinearA2}) represent semantic-sparse dimensions, whereas subspaces with larger singular values (geometrically complementary to the null space and termed row space \cite{LinearA1, LinearA2}) encode rich semantic information and are referred to as the semantic-rich subspaces.
(2) \textbf{Clipping of null space}: 
The null space typically constitutes a significant proportion of the semantic space dimensions in recommendation (\eg nearly 80\%, as shown later in Figure \ref{fig:singular_values}). 
By discarding some dimensions, we clip the null space to a suitable size, commonly set to 64 or 128 dimensions, to prepare for encoding collaborative signals.
(3) \textbf{Standardization of semantic-rich subspaces}: 
The raw semantic space is highly anisotropic \cite{WhitenBert,WhitenRec}, with non-uniform distributions of information. 
This results in imbalanced weight coefficients across the semantic-sparse subspaces, reflecting disparities in semantic information. 
To address this, we standardize these weights to ensure a more balanced representation.
(4) \textbf{Learning of ID embeddings}: ID embeddings are initialized and trained within the post-clip null space, while the language embeddings standardized in the semantic-rich subspaces remain frozen during training.
This design further ensures that the collaborative signals are injected into the redundant dimensions without interfering with the preserved semantics.
We fuse these ID embeddings and language embeddings together as the final item representations.

The proposed AlphaFuse is also model-agnostic and can adapt to various sequential recommendation paradigms (\eg SASRec \cite{SASRec}, DreamRec \cite{DreamRec}). 
Our contributions can be summarized as follows:
\begin{itemize}[leftmargin=*]
    \item We provide a new perspective on language embeddings in recommendation through SVD, partitioning the semantic space into a semantic-sparse subspace (\ie null space) and the semantic-rich subspaces (\ie row space, complementary to the null space).
    \item We propose AlphaFuse, a semantic-anchored and tuning-efficient language-guided learning strategy, which incorporates collaborative signals without compromising semantics by learning ID embeddings within the null space of language embeddings. 
    \item Extensive experiments across three datasets (\ie Movies \cite{Amazon2023}, Toys \cite{Amazon2014}, Sports \cite{Amazon2014}), including cold-start user and long-tail settings, with both discriminative and diffusion-based generative backbones, validate the effectiveness, flexibility, and efficiency of our proposed AlphaFuse.
\end{itemize}

%% file: chapters/2-prelim.tex
\section{Preliminaries}\label{sec:prelim}

Let $\mathcal{V}$ be the set of items and $\mathcal{U}$ be the set of users, where $N$ and $M$ denote the number of items and users, respectively. 
An interaction history of length $L$ is represented as $\mathbf{v}_{\le L} = \left[{v}_1, {v}_2, \ldots, {v}_{L-1}, {v}_{L}\right]$.

\vspace{5pt}
\noindent\textbf{Sequential recommendation formulation:}
Sequential recommenders offer personalized item suggestions based on users' past interactions. 
Discriminative sequential recommenders \cite{SASRec, GRURec, Caser, Bert4Rec} aim to classify the positive item $\mathbf{v}_L$ from the rest of the item set $\mathcal{V}$, conditioned on the historical interactions $\mathbf{v}_{<L}$. These methods typically achieve improved classification performance through negative sampling. 
Let $\mathbf{N}$ represent the negative samples corresponding to the historical interactions $\left(\mathbf{v}_{<L}, \mathbf{v}_L\right)$. 
Therefore, training a discriminative sequential recommendation model involves modeling a discrete probability distribution as follows:
\begin{equation}
    \label{eq:dseq_train}
    \arg\max\limits_{\Theta} \log p_{\Theta}({v}_L|\mathbf{v}_{<L})+\frac{1}{|\mathbf{v}_{n}|}\sum\limits_{v'\in\mathbf{N}}\log (1-p_{\Theta}(v'|\mathbf{v}_{<L})),
\end{equation}
where $\Theta$ is a set of trainable model parameters. Correspondingly, the inference process can be expressed as follows:
\begin{equation}
    \label{eq:dseq_infer}
    \arg\max\limits_{{v}\in\mathcal{V}} p_{\Theta}({v}|\mathbf{v}_{<L}).
\end{equation}
 As for generative sequential recommenders, each item ${v} \in \mathcal{V}$ is initially mapped to its corresponding embedding vector $\mathbf{e}\in\mathbb{R}^{d}$.
 Then, the embedding of historical interactions can be represented as $\mathbf{e}_{<L} = [\mathbf{e}_1, \mathbf{e}_2, \ldots, \mathbf{e}_{L-1}]$, with positive items represented by $\mathbf{e}_{L}$.
Generative sequential recommenders, as exemplified by DreamRec \cite{DreamRec}, aim to approximate the continuous embedding distribution:
\begin{equation}
    \label{eq:gseq_train}
    \arg\max\limits_{\Theta} \log p_{\Theta}(\mathbf{e}_L|\mathbf{e}_{<L}),
\end{equation}
where $\Theta$ denotes the trainable parameters.
Thus, the inference process is given by:
\begin{equation}
    \label{eq:gseq_infer}
     \mathbf{x}\sim p_{\Theta}(\cdot|\mathbf{e}_{<L}), \mathbf{x}\in\mathbb{R}^{d}.
\end{equation}
The generated embedding $\mathbf{x}$ typically does not belong to the actual item set $\mathcal{V}$, requiring an additional step to ground it to real items.

\vspace{5pt}
\noindent\textbf{Fundamental subspaces:} Denote the embedding matrix of all items by $\mathbf{E}\in\mathbb{R}^{N\times d}$. 
Let $\text{SVD}(\mathbf{E}^T\mathbf{E})=\mathbf{U}\mathbf{\Sigma}\mathbf{U}^T$ represent the singular value decomposition of $\mathbf{E}$, where $\mathbf{U}$ contains singular vectors $\{\mathbf{u}_i\in\mathbb{R}^d\}_{i=1}^{d}$ and $\mathbf{\Sigma}$ is a diagonal matrix with diagonal elements $\sigma_i^2$, representing the squares of the singular values sorted in descending order.
Under the singular vectors $\{\mathbf{u}_i\in\mathbb{R}^d\}_{i=1}^{d}$ as the bases of the linear space $\mathbb{R}^d$, we define the following fundamental subspaces:
\begin{itemize}[leftmargin=*]
    \item \textit{Singular Subspace:} 
    The singular subspace corresponding to a singular value $\lambda$ is defined as: $\mathbf{V}_{\lambda}=\{\mathbf{x}\in\mathbb{R}^d|\mathbf{E}\cdot\mathbf{x}=\lambda\mathbf{x}\} = \text{span}(\{\mathbf{u}_i|\sigma_i=\lambda\})$, which is spanned by all singular vectors associated with the singular value $\lambda$.
    \item \textit{Row Space:} $\mathbf{V}_{\mathbf{E}}=\text{span}(\{\mathbf{u}_i\}_{i=1}^{d}$, same as the span of rows of $\mathbf{E}$.
    \item \textit{Null Space:} The null space of $\mathbf{E}$, denoted as $\mathbf{V}_{0}$, is the singular subspace corresponding to the singular value 0: $\mathbf{V}_{0}=\text{span}(\{\mathbf{u}_i|\sigma_i=0\})$.
    Geometrically, the matrix $E$ has no projection onto the null space, meaning it is orthogonal to the null space.
\end{itemize}

%% file: chapters/3-method.tex
\section{Method}\label{sec:method}

In this section, we introduce AlphaFuse, a new language-guided learning strategy for ID embeddings that efficiently integrates semantic information with collaborative signals. 
An overview of the proposed AlphaFuse is shown in Figure \ref{figframework}.
We begin by applying singular value decomposition (SVD) \cite{LinearA1} to the language embeddings, partitioning the semantic space into singular subspaces associated with distinct singular values. These subspaces are then categorized into two groups: the semantic-sparse space (\ie null space) and semantic-rich subspaces (\ie row space, complementary to the null space).
Next, we perform targeted preprocessing to each subspace --- clipping the null space and standardizing the semantic-rich subspaces --- to optimize their suitability for various sequential recommendation frameworks. 
Finally, we learn the ID embeddings within the post-clip null space, while preserving the semantic information in the standardized semantic-rich subspaces. 
This process results in language-guided ID embeddings that are both semantic-anchored and parameter-free.

\subsection{Modeling Language Embeddings}

In real-world recommendation scenarios, items often come with textual metadata like attributes and descriptions \cite{Amazon2014,Amazon2023}. 
%
%
To capture the semantic and contextual knowledge of items, we utilize LLMs to convert the textual metadata of items into language embeddings. 
Specifically, we concatenate the textual attributes and descriptions of items, then input them into open-source large language models like Llama \cite{LLAMA}, or use public APIs like OpenAI's text-embedding-3, to obtain language embeddings. 
In this study, we employ the text-embedding-3 model provided by OpenAI\footnote{\url{https://platform.openai.com/docs/guides/embeddings}}.
Let the language embeddings be denoted as $\mathbf{E} \in \mathbb{R}^{N \times d_l}$, where $d_l$ is the dimensionality of the language embeddings. 
For each item $v \in \mathcal{V}$, its corresponding language embedding is represented as $\mathbf{e}_v$.

\subsection{Decomposition of Semantic Space} \label{sec:dsc}


Given the language embeddings $\mathbf{E}$, its rank-$r$ approximation $\mathbf{E}^{*}_r$ obtained via singular value decomposition (SVD) minimizes the Frobenius norm error $\| E - E_r \|_F^2$ \cite{LinearA1,LinearA2}, thereby preserving the semantics to the greatest extent.
To perform SVD, we first compute the mean and covariance of language embeddings $\mathbf{E}$ as follows:
\begin{equation}
    \label{eq:u_mean}
    \mathbf{\mu} = \sum\limits_{v\in\mathcal{V}} p(v)\mathbf{e}_{v},
\end{equation}
\begin{equation}
    \label{eq:u_cov}
    \mathbf{\Sigma} = \frac{1}{1-\sum\limits_{v\in\mathcal{V}} p(v)^2}\sum\limits_{v\in\mathcal{V}} p(v)(\mathbf{e}_{v}-\mathbf{u})(\mathbf{e}_{v}-\mathbf{u})^T,
\end{equation}
where $p(v)$ represents the normalized weight of item v, set uniformly across all items. We leave the exploration of $p(v)$, such as item popularity \cite{zipfian}, in future work.
Then, we perform SVD on the covariance matrix $\mathbf{\Sigma}$ of the language embeddings $\mathbf{E}$:
\begin{equation}
    \mathbf{U}\mathbf{S}\mathbf{U}^T  = \text{SVD}(\mathbf{\Sigma}).
\end{equation}
Geometrically, SVD decomposes $\mathbf{\Sigma}$ into a sequence of transformations \cite{LinearA2}:
a rotation or reflection in the semantic space (represented by $\mathbf{U}^T$), followed by scaling (represented by $\mathbf{S}$), and a subsequent rotation or reflection for recovery (represented by $\mathbf{U}$).
The scaling factors $S=\{s_i^2\in\mathbb{R}\}_{i=1}^d$, commonly known as squared singular values (or feature values), determine the weights of the bases (\ie singular vectors $\mathbf{U}=\{\mathbf{u}_i\in\mathbb{R}^{d_l}\}_{i=1}^{d}$) in the semantic space after the rotation or reflection. 
These scaling factors are crucial for understanding how the data is distributed across the space.
Based on these weights, the semantic space is partitioned into different subspaces, referred to as singular subspaces.

We rotate or reflect the semantic space through singular vectors $\mathbf{U}$ of language embeddings $\mathbf{E}$, and then decompose it into singular subspaces corresponding to distinct semantic weights (\ie squared singular values $\{s_i^2\}_{i=1}^{d_l}$). 
Each singular vector $\mathbf{u}_i$ forms a one-dimensional singular subspace $\mathbf{V}_{s_i}=\{\mathbf{x}\in\mathbb{R}^{d_l}|\mathbf{E}\cdot\mathbf{x}=s_i\mathbf{x}\} = \text{span}(\{\mathbf{u}_i\})$ corresponding to singular value $s_i$. 
Singular subspaces associated with larger singular values retain more of the projection of $\mathbf{E}$, thus preserving more semantic information. 
In contrast, singular subspaces with zero singular values, referred to as the null space, are orthogonal to $\mathbf{E}$ and contain no semantic information.

\begin{figure}[t!]
    \centering
    \includegraphics[width=0.4\textwidth]{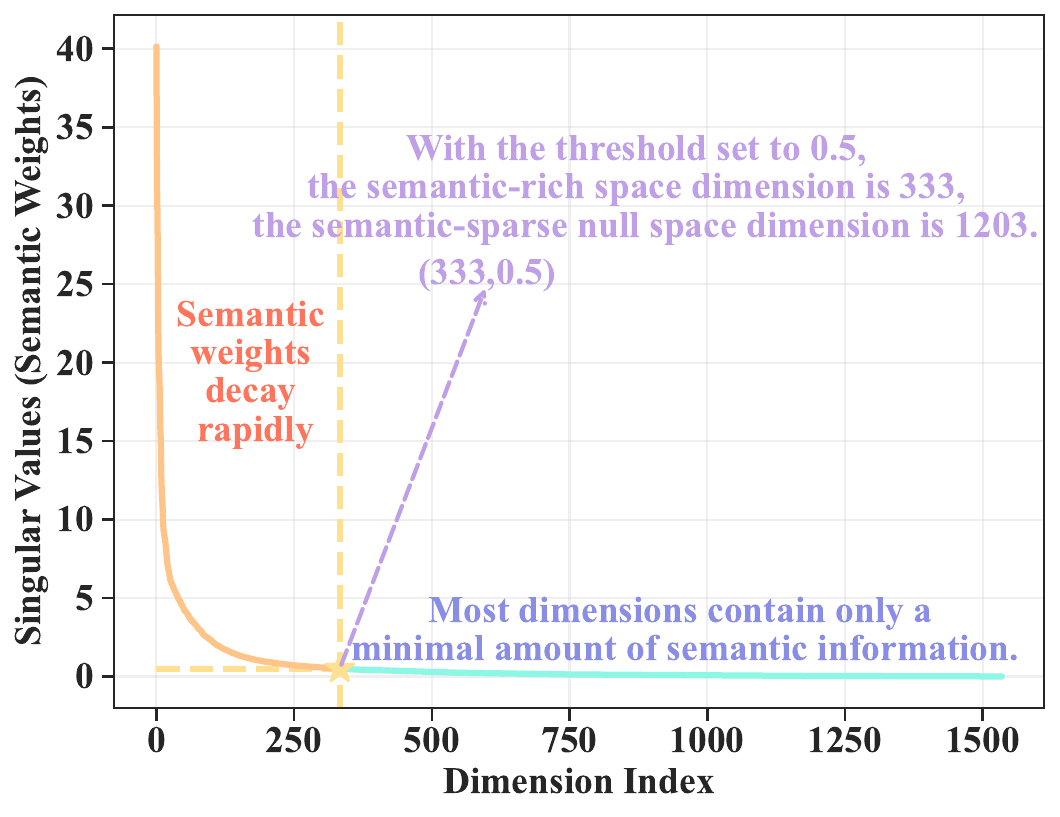}
    \vspace{-8pt}
    \caption{Normalized singular values of language embeddings for the Movies \cite{Amazon2023}. For concision, we omit plots of other datasets as they exhibit similar trends.}
    \label{fig:singular_values}
    \vspace{-5pt}
\end{figure}

\subsection{Preprocessing of Singular Subspaces} \label{sec:pss}
 
However, due to the computational errors inherent in SVD, the minimal singular values hardly reach exactly zero. To address this, we predefine a threshold and classify the subspaces with singular values below this threshold as the null space.
Alternatively, the subspaces associated with the smallest singular values may be directly designated as the null space.
The distribution of singular values (\ie semantic weights) for language embeddings $\mathbf{E}$ is illustrated in Figure \ref{fig:singular_values}.
We have two observations:
(1) \textbf{Low-rank structure of semantic space}: A large portion of the semantic information is concentrated in a small number of singular subspaces (\ie the semantic-rich subspaces), while the remaining subspaces (\ie semantic-sparse subspaces) contain negligible semantic information.
(2) \textbf{Anisotropy of semantic space}: Even among the semantic-rich subspaces, considerable variation exists in their semantic weights, reflecting an imbalance in the distribution of semantic information.
To address these challenges and facilitate the injection of collaborative signals, we propose two key preprocessing steps:
clipping the emantic-sparse space to an appropriate dimensionality and standardizing the semantic weights of the semantic-rich subspaces.

\begin{figure}[t!]
    \centering
    \includegraphics[width=0.4\textwidth]{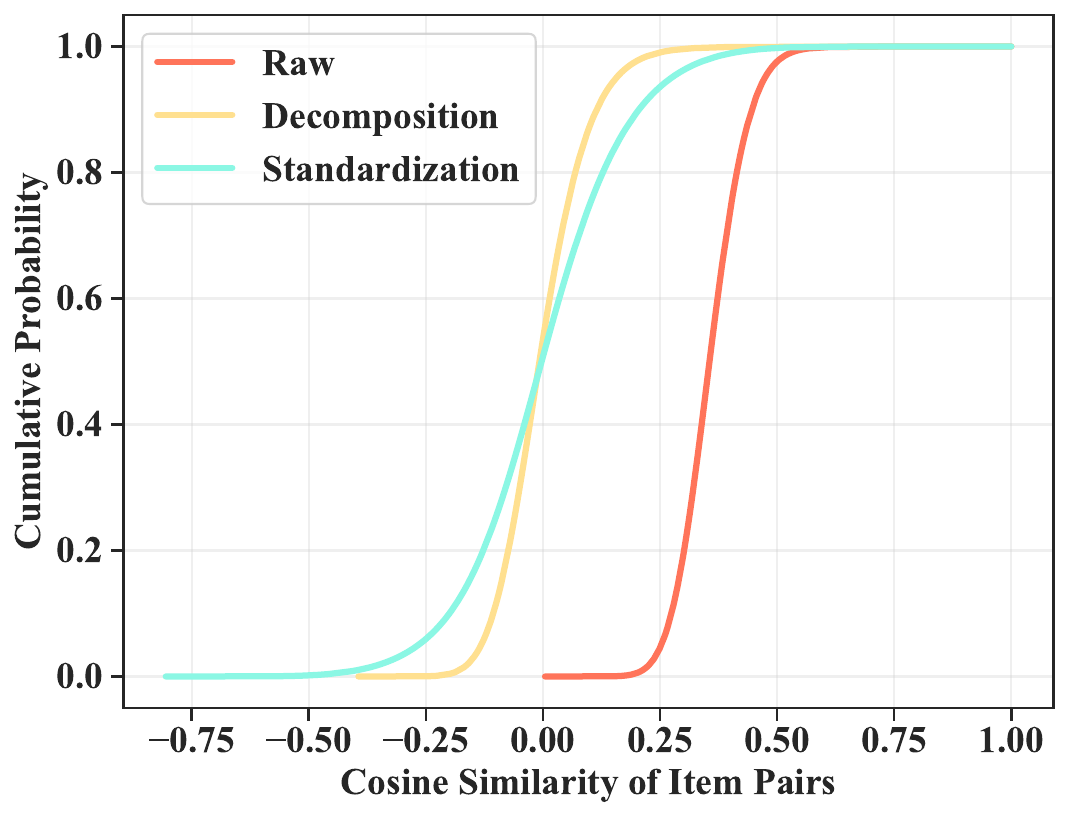}
    \vspace{-8pt}
    \caption{The cumulative distribution of cosine similarity for item pairs using language embeddings in the Movies \cite{Amazon2023}. Other datasets exhibit similar trends.}
    \label{fig:ECDF}
    \vspace{-8pt}
\end{figure}

\vspace{5pt}
\noindent\textbf{Division of Singular Subspaces}. 
By setting a threshold or directly assigning, we classify the singular subspaces $\{\mathbf{V}_{s_1},\cdots, \mathbf{V}_{s_ i},\cdots,\mathbf{V}_{s_{d_l}}\}$ into: 1) null space $\{\mathbf{V}_{s_{d_s+1}},\cdots, \mathbf{V}_{s_{d_l}}\}$, which corresponds to the semantic-sparse subspaces that contain negligible semantics, and 2) row space $\{\mathbf{V}_{s_1},\cdots, \mathbf{V}_{s_{d_{s}}}\}$, geometrically complementary to $\mathbf{V}_0$, which depicts the semantic-rich subspaces that encode meaningful semantics. 
Here, $d_s$ is the dimension of semantic-rich subspaces.

\vspace{5pt}
\noindent\textbf{Clipping of Semantic-sparse Null Space}.
Due to the low-rank structure of the semantic space in recommendation, the dimension of the null space $d_l-d_s$ is much larger than that of the semantic-rich subspaces $d_s$. 
However, for traditional discriminative sequential recommenders, the appropriate behavior space represented by ID embeddings is inherently lower-dimensional (\eg 64 or 128). 
An excessively high-dimensional null space hinders the learning of ID embeddings. 
Therefore, it is important to clip the null space to an appropriate dimensionality $d_n$ for traditional discriminative sequential recommenders \cite{SASRec,Bert4Rec}.
The clipped null space is represented by $\mathbf{V}_0$ with dimensionality $d_n$. 
In contrast, diffusion-based generative sequential recommenders are well-suited to high-dimensional space and do not require null space clipping \cite{PreferDiff}.

\vspace{5pt}
\noindent\textbf{Standardization of semantic-rich Subspaces.} 
To address the anisotropy of the raw semantic space, we standardize the bases of the semantic-rich subspaces by their semantic weights as follows:
\begin{equation}
    \mathbf{V}_{1}=\frac{1}{s_i}\mathbf{V}_{s_i}=\{\frac{1}{s_i}\mathbf{x}|\mathbf{x}\in \mathbf{V}_{s_i}\}.
\end{equation}
This standardization mitigates the disparity in semantic information across semantic-rich subspaces, ensuring more effective and distinguishable language embeddings for sequential recommendation \cite{WhitenRec}. 
As illustrated in Figure \ref{fig:ECDF}, raw language embeddings --- directly derived from a language embedding model and encoded the rich semantic space of world knowledge ---  tend to exhibit higher inner similarity. 
By decomposing the semantic space into semantic-rich and semantic-sparse subspaces, the cosine similarity within the language embeddings is reduced.
Standardizing the semantic-rich subspaces balances the disparity in semantic information, further enhancing the distinction between language embeddings and, in turn, improving the recommendation performance.


\subsection{Learning of ID embeddings}

Based on the illustrations in Sections \ref{sec:dsc} and \ref{sec:pss}, the ID embeddings, denoted as $\mathbf{E}_{\text{ID}}$, are learned within the post-clip null space $\mathbf{V}_0$, while the language embeddings remain frozen in the standardized semantic-rich subspaces $\mathbf{V}_1$.
Specifically, the language embeddings in the standardized semantic-rich subspaces $\mathbf{V}_1$ are computed as:
\begin{equation}
    \label{eq:sns}
    \mathbf{E}_{\text{language}} =(\mathbf{E}-\mathbf{\mu})\cdot\mathbf{U}[\colon,\colon d_s+d_n]\cdot\mathbf{S}^{-\frac{1}{2}}[\colon d_s+d_n,\colon d_s+d_n],
\end{equation}
where $d_s$ is the dimension of standardized semantic-rich subspaces $\mathbf{V}_1$, $d_n$ is the dimension of post-clip null space $\mathbf{V}_0$, and $\mathbf{\mu}$ represents the mean of $\mathbf{E}$.
In other words, we pre-allocate $d_n$ dimensions of language embeddings in advance to store the learned ID embeddings.
The ID embeddings $\mathbf{E}_{\text{ID}}\in\mathbb{R}^{N\times d_n}$ are initialized as either zeros or random values and updated during training. 
To extract the information from both embeddings, the final item embeddings are computed by concatenating the ID embeddings with the language embeddings, where the ID embeddings are padded to match the language embedding dimension with a zero matrix:
\begin{equation}
    \mathbf{E}_{\text{item}} =\mathbf{E}_{\text{language}}+(\mathbf{0}\in\mathbb{R}^{N\times d_s},\mathbf{E}_{\text{ID}}).
\end{equation}

It is worth noting that, in Equation \eqref{eq:sns}, we also perform standardization on the language embeddings within the null space for two main reasons: 
First, due to the dimensionality requirements of traditional discriminative recommenders (\eg $d_s+d_n=128$), we set a relatively high semantic weight threshold. 
As a result, the null space contains a small amount of semantic information, which we use to initialize the ID embeddings. 
Second, diffusion-based generative recommenders predominantly adopt Gaussian-based diffusion models (\eg DDPM \cite{DDPM,iDDPM}, DDIM \cite{DDIM}). 
In this context, standardization is essential to maintain the variance-preserving property inherent to DDPM and DDIM:
\begin{equation}
    \mathbf{E}_{\text{noise}} =\sqrt{\alpha_t}\mathbf{E}_{\text{item}}+\sqrt{1-\alpha_t}\mathbf{\epsilon}, \mathbf{\epsilon}\sim \mathcal{N}(\mathbf{0},\mathbf{I}_{d_s+d_n}),
\end{equation}
\begin{equation}
    \mathbf{E}_{\text{noise}}^T\mathbf{E}_{\text{noise}} =\alpha_t\mathbf{E}_{\text{item}}^T\mathbf{E}_{\text{item}}+(1-\alpha_t)\mathbf{I}_{d_s+d_n},
\end{equation}
\begin{align}
    \mathbf{E}_{\text{item}}^T\mathbf{E}_{\text{item}} &=\mathbf{E}_{\text{language}}^T\mathbf{E}_{\text{language}}+\mathbf{E}_{\text{ID}}^T\mathbf{E}_{\text{ID}} \\
    &=\mathbf{I}_{d_s+d_n}+\mathbf{E}_{\text{ID}}^T\mathbf{E}_{\text{ID}},
\end{align}
where $\mathcal{N}(\mathbf{0},\mathbf{I}_{d_s+d_n})$ denotes the standard Gaussian distribution, and $\mathbf{I}_{d_s+d_n}$ represents the $(d_s+d_n)$-dimensional identity matrix. 
Since $\mathbf{E}_{\text{ID}}$ is initialized by $\mathcal{N}(\mathbf{0},\mathbf{I}_{d_n})$, the variance $\mathbf{E}_{\text{noise}}^T\mathbf{E}_{\text{noise}}$ is preserved during the forward diffusion process.


%% file: chapters/4-exp.tex
\section{Experiment}\label{sec:exp}

This section presents the experimental evaluation of our AlphaFuse across multiple datasets to address the following research questions:
\begin{itemize}[leftmargin=*]
    \item \textbf{RQ1}: Does AlphaFuse outperform other language-guided methods across different sequential recommendation paradigms?
    \item \textbf{RQ2}: How effectively does AlphaFuse mitigate the cold-start user and long-tail challenges in its performance?
    \item \textbf{RQ3}: How do the dimensions of the learnable ID embeddings and the final item embeddings affect the performance of AlphaFuse?
    \item \textbf{RQ4}: What is the training and inference efficiency of AlphaFuse?
\end{itemize}

\subsection{Experimental Settings}

\subsubsection{\textbf{Datasets.}}

We evaluate AlphaFuse on three real-world datasets on Amazon:
\textbf{Movies} \cite{Amazon2023} contains movie details and user reviews from Jun 1996 to Sep 2023. 
\textbf{Toys} \cite{Amazon2014} includes user reviews and metadata for toys and games from Jun 1996 to Jul 2014.
\textbf{Sports} \cite{Amazon2014} comprises user reviews and metadata for sports and outdoor products from Jun 1996 to Jul 2014.
The statistical characteristics of the processed dataset are shown in Table \ref{stat_dataset}.

\begin{table}[t]
\vspace{-5pt}
\caption{Statistics of datasets after preprocessing.}
\label{stat_dataset}
\vspace{-8pt}
\centering
\begin{tabularx}{0.42\textwidth}{lcccc}
\toprule 
Dataset & \# users & \# items & \# Interactions & sparsity \\
\midrule
Movies & 20,515 & 44,014 & 637,157 & 00.07\%   \\
Toys & 19,412 & 11,924 &  138,444 & 00.06\%   \\
Sports & 35,598 & 18,357 & 256,598 & 00.04\%   \\
\midrule
\end{tabularx}
\vspace{-8pt}
\end{table}

\begin{table*}[ht!]
\centering
\arraybackslash
\footnotesize
\caption{Performance comparison across different backbones and methods on three datasets with cold-start user settings. Boldface indicates the highest score, while underlining denotes the second-best result among the models. }
\vspace{-5pt}
\begin{tabularx}{0.98\textwidth}{l|c|cccc|cccc|cccc}
\toprule
\multicolumn{2}{c|}{Dataset} & \multicolumn{4}{c|}{Movies} & \multicolumn{4}{c|}{Toys} & \multicolumn{4}{c}{Sports} \\
\midrule
 Backbone & Method & N@10 & M@10 & N@20 & M@20 & N@10 & M@10 & N@20 & M@20 & N@10 & M@10 & N@20 & M@20 \\
\midrule
\multirow{10}{*}{\textbf{SASRec}} 
& Base   
& 0.0338 & 0.0238 & 0.0429 & 0.0263
& 0.0255 & 0.0191 & 0.0321 & 0.0210 
& 0.0073 & 0.0049 & 0.0101 & 0.0057 
\\
& MoRec 
& 0.0154 & 0.0105 & 0.0205 & 0.0119 
& 0.0114 & 0.0069 & 0.0146 & 0.0078 
& 0.0098 & 0.0074 & 0.0109 & 0.0077 
\\
& UniSRec 
& 0.0232 & 0.0160 & 0.0303 & 0.0179 
& 0.0271 & 0.0191 & 0.0311 & 0.0202 
& 0.0071 & 0.0051 & 0.0084 & 0.0055
\\
& WhitenRec 
& 0.0168 & 0.0116 & 0.0223 & 0.0131
& 0.0258 & 0.0181 & 0.0304 & 0.0194 
& \underline{0.0115} & \underline{0.0081} & \underline{0.0141} & \underline{0.0088} 
\\
& RLMRec-Con 
& 0.0346 & 0.0244 & 0.0441 & 0.0269
& 0.0266 & 0.0185 & 0.0304 & 0.0195 
& 0.0089 & 0.0058 & 0.0107 & 0.0063
\\
& RLMRec-Gen 
& 0.0355 & 0.0252 & 0.0449 & 0.0278
& \underline{0.0303} & \underline{0.0246} & \underline{0.0347} & \underline{0.0257}
& 0.0080 & 0.0054 & 0.0102 & 0.0060 
\\
& LLMInit 
& \underline{0.0370} & \underline{0.0264} & \underline{0.0470} & \underline{0.0291}
& 0.0275 & 0.0215 & 0.0313 & 0.0225 
& 0.0083 & 0.0055 & 0.0102 & 0.0060 
\\
& LLM-ESR 
& 0.0139 & 0.0094 & 0.0192 & 0.0108
& 0.0122 & 0.0104 & 0.0153 & 0.0112 
& 0.0101 & 0.0075 & 0.0118 & 0.0079 
\\
& \cellcolor{blue!10}AlphaFuse
& \cellcolor{blue!10}\textbf{0.0459}
& \cellcolor{blue!10}\textbf{0.0324}
& \cellcolor{blue!10}\textbf{0.0574}
& \cellcolor{blue!10}\textbf{0.0355}
& \cellcolor{blue!10}\textbf{0.0339}
& \cellcolor{blue!10}\textbf{0.0287}
& \cellcolor{blue!10}\textbf{0.0376}
& \cellcolor{blue!10}\textbf{0.0297}
& \cellcolor{blue!10}\textbf{0.0137}
& \cellcolor{blue!10}\textbf{0.0098}
& \cellcolor{blue!10}\textbf{0.0158}
& \cellcolor{blue!10}\textbf{0.0104}
\\
& Best Impr. & 
\cellcolor{green!10}\textbf{+24.05\%} & \cellcolor{green!10}\textbf{+22.73\%} & 
\cellcolor{green!10}\textbf{+22.13\%} & 
\cellcolor{green!10}\textbf{+21.99\%} & 
\cellcolor{green!10}\textbf{+11.88\%} & \cellcolor{green!10}\textbf{+16.67\%} & 
\cellcolor{green!10}\textbf{+8.36\%} & 
\cellcolor{green!10}\textbf{+15.56\%} &
\cellcolor{green!10}\textbf{+19.13\%} & \cellcolor{green!10}\textbf{+20.99\%} & 
\cellcolor{green!10}\textbf{+12.06\%} & 
\cellcolor{green!10}\textbf{+18.18\%} \\
\midrule
\multirow{11}{*}{\textbf{DreamRec}} 
& Base 
& 0.0016 & 0.0013 & 0.0018 & 0.0014 
&  \underline{0.0383} &  \underline{0.0333}
&  \underline{0.0392} &  \underline{0.0336}
&  \underline{0.0158} &  \underline{0.0132} 
&  \underline{0.0170} &  \underline{0.0135} 
\\
& iDreamRec 
& \underline{0.0226} & \underline{0.0180}
& \underline{0.0262} & \underline{0.0189}
& 0.0350 & 0.0301 & 0.0373 & 0.0307
& 0.0141 & 0.0119 & 0.0155 & 0.0123
\\
& MoRec 
& 0.0002 & 0.0002 & 0.0003 & 0.0002
& 0.0030 & 0.0026 & 0.0034 & 0.0027 
& 0.0012 & 0.0010 & 0.0017 & 0.0012 
\\
& UniSRec 
& 0.0021 & 0.0014 & 0.0030 &  0.0017
& 0.0014 & 0.0008 & 0.0022 & 0.0010 
& 0.0004 & 0.0002 & 0.0008 & 0.0003
\\
& WhitenRec 
& 0.0007 & 0.0006 & 0.0008 & 0.0006  
& 0.0029 & 0.0021 & 0.0034 & 0.0022 
& 0.0026 & 0.0019 & 0.0030 & 0.0021
\\
& RLMRec
& 0.0016 & 0.0013 & 0.0019 & 0.0014
& 0.0376 & 0.0321 & 0.0388 & 0.0325
& \underline{0.0160} & \underline{0.0135}
& \underline{0.0172} & \underline{0.0138}
\\
& LLMInit 
& 0.0082 & 0.0056 & 0.0113 & 0.0065 
& 0.0198 & 0.0179 & 0.0214 & 0.0184
& 0.0075 & 0.0065 & 0.0086 & 0.0068 
\\
& LLM-ESR 
& 0.0007 & 0.0004 & 0.0010 & 0.0005
& 0.0073 & 0.0061 & 0.0090 & 0.0066
& 0.0045 & 0.0037 & 0.0048 & 0.0037 
\\
& \cellcolor{blue!10}AlphaFuse
& \cellcolor{blue!10}\textbf{0.0246}
& \cellcolor{blue!10}\textbf{0.0201}
& \cellcolor{blue!10}\textbf{0.0279}
& \cellcolor{blue!10}\textbf{0.0209}
& \cellcolor{blue!10}\textbf{0.0408}
& \cellcolor{blue!10}\textbf{0.0348}
& \cellcolor{blue!10}\textbf{0.0425}
& \cellcolor{blue!10}\textbf{0.0353}
& \cellcolor{blue!10}\textbf{0.0165}
& \cellcolor{blue!10}\textbf{0.0139}
& \cellcolor{blue!10}\textbf{0.0174}
& \cellcolor{blue!10}\textbf{0.0142}
\\
& Best Impr. 
& \cellcolor{green!10}\textbf{+8.85\%} & \cellcolor{green!10}\textbf{+11.67\%}
& \cellcolor{green!10}\textbf{+6.49\%} & \cellcolor{green!10}\textbf{+10.58\%}
& \cellcolor{green!10}\textbf{+6.53\%} & \cellcolor{green!10}\textbf{+4.50\%}
& \cellcolor{green!10}\textbf{+8.42\%} & \cellcolor{green!10}\textbf{+5.06\%}
& \cellcolor{green!10}\textbf{+3.13\%} & \cellcolor{green!10}\textbf{+2.96\%}
& \cellcolor{green!10}\textbf{+1.16\%} & \cellcolor{green!10}\textbf{+2.90\%}
\\
\bottomrule
\end{tabularx}
\label{tab:cold-start}
\end{table*}

\subsubsection{\textbf{Experimental Settings.}}
For a comprehensive comparison, we adopt two settings: leave-one-out and cold-start user settings.
\textbf{Leave-One-Out Setting \cite{SASRec}}: In the leave-one-out setting, the last item in a user's interaction sequence is designated as the test item, the second-to-last item is used for validation, and the remaining items are used for training. 
This setting is commonly used in sequential recommendation tasks. 
However, the leave-one-out setting has a clear limitation, \ie it requires all users to appear in the training dataset, making it unsuitable for evaluating cold-start users, who have limited historical interactions.
\textbf{Cold-Start User Setting \cite{DreamRec}}: To address the limitation, we also adopt the cold-start user setting. 
In this case, users are chronologically sorted based on the time of their last interaction. 
The most recent users are assigned to the test set, the next most recent users to the validation set, and the remaining users are included in the training set. 
In our experiments, we split the data into training, validation, and test sets with an 8:1:1 ratio, while ensuring that the last 10 interactions are preserved as the historical sequence for each user.

\subsubsection{\textbf{Backbones.}}
To validate the flexibility of AlphaFuse, we selected two representative sequential recommendation models: SASRec \cite{SASRec} and DreamRec \cite{DreamRec}. 
The former represents the traditional discriminative paradigm, while the latter follows the generative paradigm based on diffusion models. 
For SASRec, we employed InfoNCE loss \cite{InfoNCE} with 64 negative samples. 
For DreamRec, we used the well-known MSE \cite{DDPM} for Gaussian-based diffusion models.

\subsubsection{\textbf{Baselines.}}
We compare AlphaFuse with a range of existing language-guided ID embedding learning strategies, including semantic reconstruction method (RLMRec), semantic initialization method (LLMInit), adaptive projection methods (MoRec, UniSRec, WhitenRec), and other baselines (LLM-ESR, iDreamRec):
\textbf{RLMRec} \cite{RLMRec} 
originally designed for collaborative filtering, we adapt it for sequential recommendation:
\textbf{RLMRec-Gen} maps language embeddings into behavior space via a 2-layer MLP, using similarity differences to ID embeddings as a regularization term in the loss.
\textbf{RLMRec-Con} maps ID embeddings into semantic space via a 2-layer MLP and computes similarity differences to language embeddings instead.
\textbf{LLMInit} \cite{LLM2BERT4Rec,SAID,elephant} uses language embeddings to initialize ID embeddings. To align with behavior space, we apply PCA to reduce the dimensionality of the language embeddings. 
\textbf{MoRec} \cite{MoRec} uses a pre-trained modal encoder followed by a dense layer for dimension transformation as the item encoder.
\textbf{UniSRec} \cite{UniSRec} processes language embeddings through a mixture-of-experts enhanced adaptor to produce final item embeddings.
\textbf{WhitenRec} \cite{WhitenRec} uses a whitening transformation \cite{WhitenBert} to address the issue of high average similarity in language embeddings. 
The whitened language embeddings are then passed through an MLP adapter to obtain the final item embeddings.
\textbf{LLM-ESR} \cite{LLMESR} proposes a dual-view modeling approach that incorporates semantic initialization and adaptive projection to address the long-tail user and item problem.
\textbf{iDreamRec} \cite{iDreamRec} proposes using diffusion models to model consistent language embeddings for recommendation, while employing a linear transformation to achieve variance-preserving.

\subsubsection{\textbf{Implementation Details.}} 

\noindent\textbf{Dataset settings}:
For the leave-one-out setting, we replicate the results from LLM-ESR \cite{LLMESR}. 
For AlphaFuse, we strictly follow the dataset splitting and code from LLM-ESR \footnote{\url{https://github.com/liuqidong07/LLM-ESR}}, only adjusting the learning rate to ${0.001, 0.0001, 0.00001}$.
For the cold-start user setting, 
we concatenate the textual attributes and descriptions of items and employ the text-embedding-3 provided by OpenAI\footnote{\url{https://platform.openai.com/docs/guides/embeddings}} to derive language embeddings. 
\begin{table*}[ht!]
\centering
\caption{Performance comparison across different methods on three datasets with long-tail settings. }
\vspace{-5pt}
\begin{tabularx}{0.98\textwidth}{cl|cc|cccc|cccc}
\toprule
\multirow{2}{*}{\textbf{Dataset}} & \multirow{2}{*}{\textbf{Model}} & \multicolumn{2}{c|}{\textbf{Overall}} & \multicolumn{2}{c}{\textbf{Tail Item}} & \multicolumn{2}{c|}{\textbf{Head Item}} & \multicolumn{2}{c}{\textbf{Tail User}} & \multicolumn{2}{c}{\textbf{Head User}} \\
\cmidrule{3-12}
  & & R@10 & N@10 & R@10 & N@10 & R@10 & N@10 & R@10 & N@10 & R@10 & N@10  \\
\midrule
\multirow{4}{*}{\textbf{Yelp}} 
& SASRec &
 0.5940 & 0.3597 &
 0.1142 & 0.0495 & 0.7353 & 0.4511 &
 0.5893 & 0.3578 & 0.6122 & 0.3672 \\
& -LLM-ESR &
 \textbf{0.6673} & 0.4208 &
 \textbf{0.1893} & \textbf{0.0845} & \textbf{0.8080} & 0.5199 &
 \textbf{0.6685} & 0.4229 & \textbf{0.6627} & 0.4128 \\
&\cellcolor{blue!10}-\textbf{AlphaFuse}
& \cellcolor{blue!10}0.6631
&\cellcolor{blue!10}\textbf{0.4219}
&\cellcolor{blue!10} 0.1815
& \cellcolor{blue!10}0.0775
& \cellcolor{blue!10}0.8048
& \cellcolor{blue!10}\textbf{0.5232}
& \cellcolor{blue!10}0.6617
& \cellcolor{blue!10}\textbf{0.4239}
& \cellcolor{blue!10}0.6585
& \cellcolor{blue!10}\textbf{0.4141}
\\
& Best Impr. & 
 \cellcolor{red!10}-0.63\% & \cellcolor{green!10}\textbf{+0.26\%} & 
 \cellcolor{red!10}-4.12\% & \cellcolor{red!10}-8.28\% & \cellcolor{red!10}-0.40\% & \cellcolor{green!10}\textbf{+0.63\%} & \cellcolor{red!10}-1.02\% & \cellcolor{green!10}\textbf{+0.24\%} & \cellcolor{red!10}-0.63\% & \cellcolor{green!10}\textbf{+0.31\%} \\
\midrule
\multirow{4}{*}{\textbf{Fashion}} 
& SASRec &
 0.4956 & 0.4429 &
 0.0454 & 0.0235 & 0.6748 & 0.6099 &
 0.3967 & 0.3390 & 0.6239 & 0.5777 \\
& -LLM-ESR &
 0.5619 & 0.4743 &
 0.1095 & 0.0520 & \textbf{0.7420} & 0.6424 &
 0.4811 & 0.3769 & 0.6668 & 0.6005 \\
& \cellcolor{blue!10}-\textbf{AlphaFuse}
& \cellcolor{blue!10}\textbf{0.6008}
& \cellcolor{blue!10}\textbf{0.5103}
& \cellcolor{blue!10}\textbf{0.2601}
& \cellcolor{blue!10}\textbf{0.1646}
&\cellcolor{blue!10} 0.7364
& \cellcolor{blue!10}\textbf{0.6479}
& \cellcolor{blue!10}\textbf{0.5352}
& \cellcolor{blue!10}\textbf{0.4276}
& \cellcolor{blue!10}\textbf{0.6860}
& \cellcolor{blue!10}\textbf{0.6175}
\\
& Best Impr. & 
 \cellcolor{green!10}\textbf{+6.92\%} & \cellcolor{green!10}\textbf{+7.59\%} & \cellcolor{green!10}\textbf{+137.53\%} & \cellcolor{green!10}\textbf{+216.54\%} & 
 \cellcolor{red!10}-0.75\% & \cellcolor{green!10}\textbf{+0.86\%} & \cellcolor{green!10}\textbf{+11.25\%} & \cellcolor{green!10}\textbf{+13.45\%}& \cellcolor{green!10}\textbf{+2.88\%} & \cellcolor{green!10}\textbf{+2.83\%} \\
\midrule
\multirow{4}{*}{\textbf{Beauty}} 
& SASRec &
 0.4388 & 0.3030 &
 0.0870 & 0.0649 & 0.5227 & 0.3598 &
 0.4270 & 0.2941 & 0.4926 & 0.3438 \\
& -LLM-ESR &
 0.5672 & 0.3713 & 
 \textbf{0.2257} & \textbf{0.1108} & 0.6486 & 0.4334 &
 0.5581 & 0.3643 & 0.6087 & 0.4032 \\
& \cellcolor{blue!10}{-\textbf{AlphaFuse}} 
& \cellcolor{blue!10}{\textbf{0.5793}}
& \cellcolor{blue!10}{\textbf{0.4046}}
& \cellcolor{blue!10}{0.1625}
& \cellcolor{blue!10}{0.1006}
& \cellcolor{blue!10}\textbf{0.6787}
& \cellcolor{blue!10}\textbf{0.4771}
& \cellcolor{blue!10}\textbf{0.5692}
& \cellcolor{blue!10}\textbf{0.3984}
& \cellcolor{blue!10}\textbf{0.6258}
& \cellcolor{blue!10}\textbf{0.4326}
\\
& Best Impr. & 
\cellcolor{green!10}\textbf{+2.13\%} & 
\cellcolor{green!10}\textbf{+8.97\%} & 
 \cellcolor{red!10}-28.00\% &  \cellcolor{red!10}-9.21\% & \cellcolor{red!10}\textbf{-4.64\%} & 
\cellcolor{green!10}\textbf{+10.08\%} & 
\cellcolor{green!10}\textbf{+1.99\%} & 
\cellcolor{green!10}\textbf{+9.36\%} & 
\cellcolor{green!10}\textbf{+2.81\%}& 
\cellcolor{green!10}\textbf{+7.29\%}\\
\bottomrule
\end{tabularx}
\label{tab:long-tail}
\end{table*}
\noindent\textbf{SASRec backbone}:
We fix the dimension of the post-clip semantic space, \ie the final item embeddings to 128 for all methods, and ID embeddings to 64 for AlphaFuse and LLM-ESR. 
We employ InfoNCE loss \cite{InfoNCE} with 64 negative samples. 
For baselines that use additional loss functions, we fix their coefficients to 0.1 or 1, as suggested in their original papers \cite{RLMRec, LLMESR}. 
\noindent\textbf{DreamRec backbone}:
We do not clip the semantic space and maintain the original dimension of the language embeddings.
Therefore, RLMRec-Con and RLMRec-Gen are similar.
For LLM-ESR, we allocate equal dimensions to both the semantic-view and the collaborative-view.
We employ the MSE loss \cite{DDPM} without any negative sample.
For baselines that use additional loss functions, we fix their coefficients to 0.1 or 1, as per the original papers \cite{RLMRec, LLMESR}.
For AlphaFuse, we select the null space by setting thresholds from ${0.001, 0.01, 0.1, 0.25, 0.5}$, drop the original values in the null space, and replace them with $\mathbf{E}_{\text{ID}} \in \mathbb{R}^{N \times d_n}$, initialized with a standard Gaussian distribution.
\noindent\textbf{Training settings}: 
We implement all models with Python 3.7 and PyTorch 1.12.1 in Nvidia GeForce RTX 3090.
During training, all methods are trained with a fixed batch size of 256 and a learning rate from $\{0.01,0.001,0.0001,0.00001\}$ using the Adam optimizer. 
We also adopt the early stop technique based on the model's performance on the validation set.
To ensure reproducibility, we fix all random seeds to 22, a random number.
\noindent\textbf{Evaluation Protocols and Metrics}:
To ensure a comprehensive evaluation and mitigate potential biases, we adopt the all-rank protocol \cite{LightGCN, SGL, RLMRec}, which evaluates recommendations across all items.
We utilize two widely adopted ranking-based metrics: \textit{Normalized Discounted Cumulative Gain} (\textbf{N$@K$}) and \textit{Mean Reciprocal Rank} (\textbf{M$@K$}), to measure the effectiveness of models.

\subsection{Performance Comparison (RQ1 - RQ2)}
To validate the effectiveness and flexibility of our proposed AlphaFuse, we present the results under both the cold-start user setting and the leave-one-out setting in Table \ref{tab:cold-start} and Table \ref{tab:long-tail}, respectively. 
Table \ref{tab:cold-start} compares AlphaFuse with other language-guided methods in the cold-start user setting, using both discriminative and generative sequential recommendation backbones. 
In the leave-one-out setting, we replicate the results from LLM-ESR \cite{LLMESR}, where LLM-ESR outperformed baselines such as CITIES \cite{CITIES}, MELT \cite{MELT}, RLMRec \cite{RLMRec}, and LLMInit \cite{LLM2BERT4Rec, SAID, elephant}, effectively addressing long-tail user and item problem.
We present the comparison between AlphaFuse and LLM-ESR under SASRec backbone in Table \ref{tab:long-tail}.

\vspace{5pt}
\noindent\textbf{Overall Comparison.} 
From the results in Table \ref{tab:cold-start} and Table \ref{tab:long-tail}, we observe that AlphaFuse consistently outperforms other methods, regardless of the sequential recommendation paradigm—whether discriminative or generative—and across both cold-start user and leave-one-out settings. Notably, under the DreamRec backbone, many language-guided methods exhibit limited performance, whereas AlphaFuse maintains strong results. We attribute AlphaFuse's effectiveness and flexibility to its ability to seamlessly integrate semantic information with collaborative signals, thereby improving representation performance.

\vspace{5pt}
\noindent\textbf{Cold-start User Comparison.}
From the results in Table \ref{tab:cold-start}, AlphaFuse demonstrates superior recommendation performance for cold-start users. 
In general, we observe that the baselines do not employ additional modules during inference (\ie RLMRec, LLMInit, and AlphaFuse) perform better in addressing the cold-start user problem. 
This could be due to the limited generalization ability of the extra modules, which are specifically trained on the training dataset. 
However, both RLMRec and LLMInit rely on language embeddings solely to assist in training ID embeddings, which leads to underutilization of the semantic potential in the language embeddings. 
In contrast, AlphaFuse not only leverages language embeddings to define the null space for ID embeddings but also fuses them directly into the final item embeddings. 
This approach enables AlphaFuse to fully exploit the semantic information contained in the language embeddings, resulting in better performance for cold-start users.

\vspace{5pt}
\noindent\textbf{Long-tail Item and User Comparison.}
LLM-ESR \cite{LLMESR} is a language-guided method specifically designed to address the long-tail problem \cite{CITIES}. 
We compare AlphaFuse with LLM-ESR under its dataset and code framework.
From the results in Table \ref{tab:long-tail}, AlphaFuse also demonstrates competitive performance in addressing the long-tail user and item problem. 
Specifically, for long-tail users and items, both LLM-ESR and AlphaFuse show improvements over the naive SASRec. 
For long-tail users and head users/items, AlphaFuse generally outperforms LLM-ESR. 
However, for long-tail items, LLM-ESR performs better than AlphaFuse on the Beauty and Yelp datasets, while AlphaFuse significantly outperforms LLM-ESR on the Fashion dataset. Overall, AlphaFuse proves to be effective in tackling the long-tail problem, particularly considering that it does not introduce additional trainable modules and maintains training and inference efficiency similar to the naive SASRec.

\begin{figure*}[t!]
    \begin{subfigure}{0.325\linewidth}
        \centering
        \includegraphics[width=\textwidth]{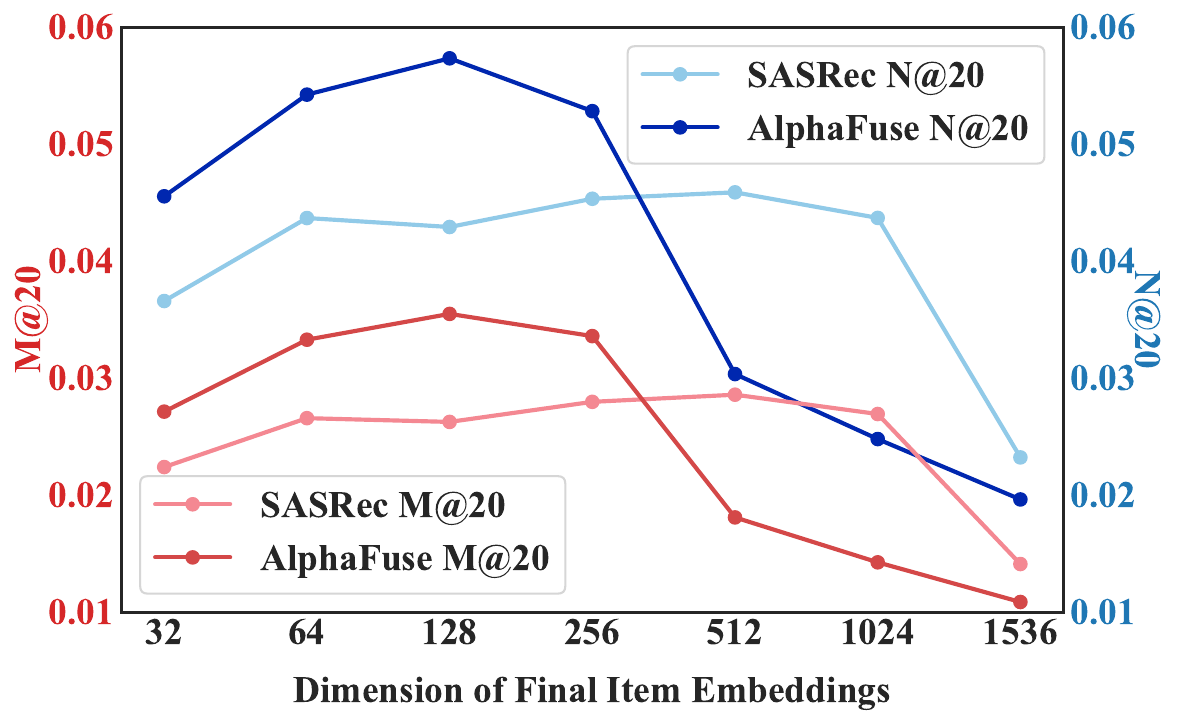}
        \vspace{-10pt}
        \caption{Dimension of Item embeddings.}
        \label{fig:dim_item_embs}
    \end{subfigure}
    \begin{subfigure}{0.325\linewidth}
        \centering
        \includegraphics[width=\textwidth]{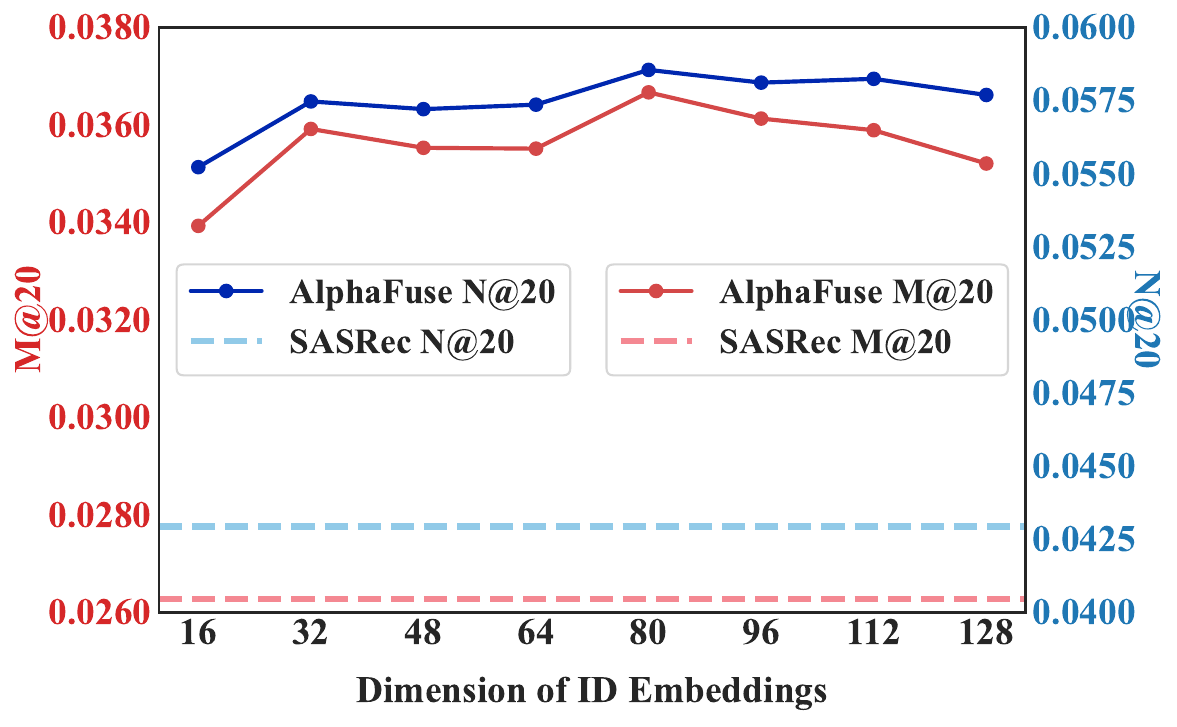}
        \vspace{-10pt}
        \caption{Dimension of ID embeddings.}
        \label{fig:dim_ID_embs}
    \end{subfigure}
    \begin{subfigure}{0.325\linewidth}
        \centering
        \includegraphics[width=\textwidth]{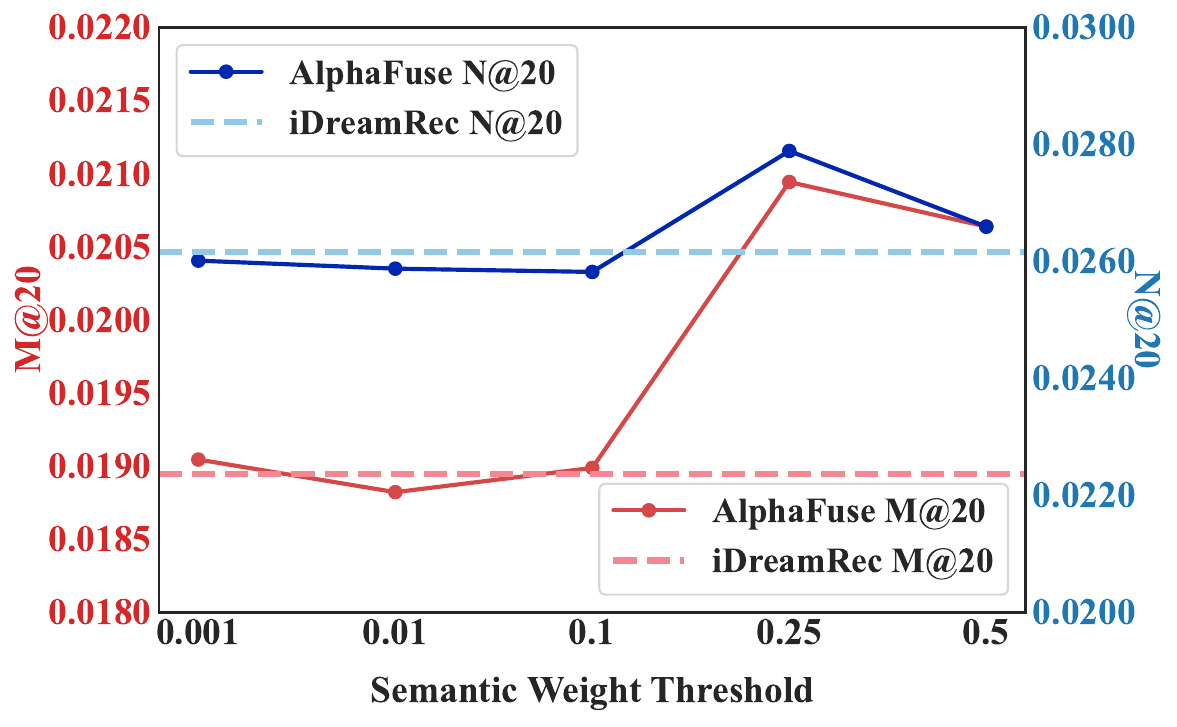}
        \vspace{-10pt}
        \caption{Threshold of Semantic Weights.}
        \label{fig:thres}
    \end{subfigure}
    \vspace{-10pt}
    \caption{Hyperparameter experiments for AlphaFuse: the left subfigures analyze the impact of final item and ID embeddings on SASRec-based AlphaFuse, while the right one evaluates the effect of semantic weight threshold on DreamRec-based AlphaFuse.}
    \vspace{-10pt}
    \label{fig:param}
\end{figure*}    

\begin{table}[t!]
\centering
\caption{The ablation study with SASRec as the backbone.}
\vspace{-5pt}
\begin{tabularx}{0.47\textwidth}{cl|cccc}
\toprule
\multirow{2}{*}{\textbf{Dataset}} & \multirow{2}{*}{\textbf{Model}} & \multicolumn{4}{c}{\textbf{SASRec Backbone}} \\
\cmidrule{3-6}
  & & N@10 & M@10 & N@20 & M@20\\
\midrule

\multirow{4}{*}{\textbf{Movies}}
& \cellcolor{blue!10}\textbf{AlphaFuse}
& \cellcolor{blue!10}0.0459
& \cellcolor{blue!10}0.0324 
& \cellcolor{blue!10}0.0574 
& \cellcolor{blue!10} 0.0355 \\
& \cellcolor{red!10}-w/o Frozen. 
& \cellcolor{red!10} 0.0350
& \cellcolor{red!10} 0.0249
& \cellcolor{red!10} 0.0444
& \cellcolor{red!10} 0.0274 \\
& \cellcolor{red!10}-w/o Clip.
& \cellcolor{red!10} 0.0148  
& \cellcolor{red!10} 0.0096  
& \cellcolor{red!10} 0.0197
& \cellcolor{red!10} 0.0109
\\
& \cellcolor{red!10}-w/o Stand. 
&  \cellcolor{red!10}0.0354
&  \cellcolor{red!10}0.0250
&  \cellcolor{red!10}0.0450
&  \cellcolor{red!10}0.0276\\
\bottomrule
\end{tabularx}
\label{tab:SASRec_abla}
\vspace{-8pt}
\end{table}

\begin{table}[t!]
\centering
\caption{The ablation study with DreamRec as the backbone.}
\vspace{-5pt}
\begin{tabularx}{0.47\textwidth}{cl|cccc}
\toprule
\multirow{2}{*}{\textbf{Dataset}} & \multirow{2}{*}{\textbf{Model}} & \multicolumn{4}{c}{\textbf{DreamRec Backbone}} \\
\cmidrule{3-6}
  & & N@10 & M@10 & N@20 & M@20\\
\midrule 
\multirow{4}{*}{\textbf{Movies}}
& \cellcolor{blue!10}\textbf{AlphaFuse}
& \cellcolor{blue!10}0.0246
& \cellcolor{blue!10}0.0201
& \cellcolor{blue!10}0.0279
& \cellcolor{blue!10}0.0209 \\
& \cellcolor{red!10}-w/o Decom. 
& \cellcolor{red!10}0.0103
& \cellcolor{red!10}0.0084
& \cellcolor{red!10}0.0120
& \cellcolor{red!10}0.0089
\\
& \cellcolor{red!10}-w/o Frozen.
& \cellcolor{red!10}0.0214
& \cellcolor{red!10}0.0177
& \cellcolor{red!10}0.0242
& \cellcolor{red!10}0.0185
\\
& \cellcolor{red!10}-w/o Stand. 
& \cellcolor{red!10}0.0114 
& \cellcolor{red!10}0.0089
& \cellcolor{red!10}0.0140
& \cellcolor{red!10}0.0095
\\
\bottomrule
\end{tabularx}
\vspace{-5pt}
\label{tab:DreamRec_abla}
\end{table}

\begin{table}[t]
\caption{Comparison of efficiency through the number of trainable parameters and GFLOPs during inference.}
\label{tab:efficiency}
\centering
\vspace{-5pt}
\begin{tabular}{lcc}
\toprule 
Models &  \# Trainable Parameters & Inference GFLOPs \\
\midrule
UniSRec   & 1.69M &  4.24  \\
LLM-ESR   & 4.10M &  3.34  \\
MoRec     & 0.28M &  0.72  \\
WhitenRec & 0.28M &  0.72  \\
RLMRec    & 7.10M &  0.22  \\
LLMInit   & 5.72M &  0.22  \\
SASRec    & 5.72M &  0.22  \\
\cellcolor{blue!10}AlphaFuse & \cellcolor{blue!10}2.90M &  \cellcolor{blue!10}0.22  \\
\midrule
\end{tabular}
\end{table}

\subsection{Ablation Study (RQ3)}

As presented in Table \ref{tab:SASRec_abla}, we perform a thorough analysis and evaluation of each key component within the SASRec-based AlphaFuse to assess their individual significance. 
The ablation study is conducted using the following three variations: 
(1) AlphaFuse-w/o-Frozen, Awhere the language embeddings in the standardized semantic-rich subspaces are trainable rather than frozen;
(2) AlphaFuse-w/o-Clip, which does not clip the null space dimensions; 
and (3) AlphaFuse-w/o-Stand, which omits the standardization of the semantic-rich subspaces. 
The results clearly demonstrate that each component of the SASRec-based AlphaFuse plays an indispensable role in its overall performance.
Similarly, as shown in Table \ref{tab:DreamRec_abla}, we systematically analyze and evaluate each individual component of the DreamRec-based AlphaFuse to understand their respective impacts and significance. 
The ablation study is performed using the following three variations: 
(1) AlphaFuse-w/o-Decom, which directly utilizes the original language embeddings and learns ID embeddings in the raw semantic space; 
(2) AlphaFuse-w/o-Frozen, where the language embeddings in the standardized semantic-rich subspaces are trainable rather than frozen; 
and (3) AlphaFuse-w/o-Stand, which does not standardize the semantic-rich subspaces. 
These findings further underscore that each component of the DreamRec-based AlphaFuse is equally critical to its overall effectiveness.

\subsection{Hyper-parameter Analysis (RQ3)}

To investigate the effects of the dimensions of the learnable ID embeddings and the final item embeddings in AlphaFuse, we present the performance trends in Figure \ref{fig:param} for the Movies dataset. 
For the final item embeddings, we fix half of the dimension to match the dimension of the learned ID embeddings, as shown in Subfigure \ref{fig:dim_item_embs}. 
When the dimension is below 512, AlphaFuse effectively integrates semantic information with collaborative signals, resulting in consistent performance improvements. 
However, when the dimension becomes too large, the model is constrained by the null space of the language embeddings, which increases the difficulty of learning the ID embeddings and ultimately leads to performance degradation. 
This underscores the importance of clipping the null space.
For the SASRec-based AlphaFuse, we further fix the dimension of the final item embeddings at 128 and adjust the dimension of the learned ID embeddings. 
The results, presented in Subfigure \ref{fig:dim_ID_embs}, demonstrate that AlphaFuse is robust to variations in the ID embedding dimension. 
For the DreamRec-based AlphaFuse, we fix the final item embeddings dimension to align with the original language embeddings dimension and apply different semantic weights (\ie singular values, whose distribution is shown in Figure \ref{fig:singular_values}) to determine the ID embedding dimension. 
The results, shown in Subfigure \ref{fig:thres}, reveal that once an appropriate threshold is established, AlphaFuse achieves significant performance improvements.

\subsection{Efficiency Analysis (RQ4)}

As shown in Table \ref{tab:efficiency}, we present the number of trainable parameters and inference GFLOPs for all methods on the Movies \cite{Amazon2023} using the SASRec backbone. 
The adaptive projection methods, which does not train ID embeddings and relies on limited collaborative signals, has fewer parameters but demonstrates suboptimal overall performance. 
In contrast, AlphaFuse not only reduces the number of parameters but also minimizes inference cost, resulting in both higher efficiency and effectiveness. 
Therefore, AlphaFuse proves to be both a highly effective and efficient solution for language-guided sequential recommendation.

%% file: chapters/5-relawork.tex
\section{Related Work}\label{sec:rela}

\vspace{5pt}
\noindent\textbf{Sequential Recommendation} has gained significant attention due to its ability to predict the next item of interest based on the sequence of items a user has interacted with \cite{Recformer, Tiger, LLMESR}. 
Early approaches primarily focused on improving the efficiency of network architectures within the traditional discriminative paradigm, with notable methods such as GRU4Rec \cite{GRURec}, Caser \cite{Caser}, SASRec \cite{SASRec}, and Bert4Rec \cite{Bert4Rec}.
Recent advancements in diffusion models, have introduced a new diffusion-based generative paradigm that contrasts with these traditional discriminative approaches. 
DreamRec \cite{DreamRec} reformulates the sequential recommendation task by generating oracle item embeddings conditioned on the user's interaction history.
Building upon this foundation, subsequent research has expanded the diffusion-based generative framework in several directions: 1) the introduction of new generative models, such as Schrödinger Bridges \cite{bridge} and discrete diffusion models \cite{DDM}; 
2) the adaptation of recommendation scenarios, including the integration of negative sampling in loss functions to capture user preferences \cite{PreferDiff}; 
and 3) the proposal of novel architectures and tasks, such as multi-interest enhanced modules \cite{DimeRec} and intention guidance \cite{iDreamRec}.

\vspace{5pt}
\noindent\textbf{Language-guided Recommenders} leverage the rich item attributes and user reviews inherent in real-world recommendation scenarios \cite{Amazon2014, Amazon2023}, which contain valuable semantic information.
With the development of pre-trained language models (e.g., BERT \cite{BERT}, Llama \cite{LLAMA}, and GPT-4 \cite{GPT4}), research has progressively explored leveraging language embeddings to enhance sequential recommendation. 
Broadly, these efforts can be classified into three categories:
\textbf{Semantic Assistance}:
Approaches like KAR \cite{KAR} use language embeddings as additional input features to enhance the learning of ID embeddings. RLMRec \cite{RLMRec} treats language embeddings as reconstruction targets to guide the learning of ID embeddings.
\textbf{Semantic Initialization}:
These works \cite{LLM2BERT4Rec, SAID, elephant} leverage language embeddings to initialize ID embeddings, setting a strong foundation for further refinement during training.
\textbf{Adaptive Projection}:
These methods pass language embeddings through a trainable adaptor to output ID embeddings, thus mapping the semantic space into the behavior space. 
For instance, MoRec \cite{MoRec} proposes an MLP adaptor for dimensional transformation, while UniSRec \cite{UniSRec} introduces a mixture-of-experts-based adaptor to enhance cross-domain recommendation transferability. 
WhitenRec \cite{WhitenRec} improves the distinguishability of language embeddings by applying whitening operations, and AlphaRec \cite{AlphaRec} utilizes an MLP adaptor to establish an isomorphic mapping between the semantic space and the behavior space, enabling user intention guidance.
Beyond these, LLM-ESR \cite{LLMESR} introduces a dual-view modeling approach that combines both semantic initialization and adaptive projection to address the long-tail user and item problem.
Although effective, these methods either introduce additional trainable modules or indirectly leverage language embeddings, making it difficult to address the degradation problem of mapping high-dimensional semantic space to the low-dimensional behavior space.

\vspace{5pt}
\noindent\textbf{Null Space} refers to the subspace orthogonal to a given matrix. 
The concept of Null Space has found diverse applications across multiple domains in AI. 
Specifically, O-LoRA \cite{O-LoRA} introduces a simple and efficient method for continual learning in language models, successfully mitigating catastrophic forgetting by learning new tasks in the null space. 
In the realm of large model safety fine-tuning, substantial evidence \cite{SafetySpotlight} suggests that safety fine-tuning techniques (\eg supervised safety fine-tuning, direct preference optimization, and unlearning) result in minimal changes to MLP weights, ensuring that unsafe inputs are effectively aligned into the null space of these weights. Furthermore, AlphaEdit \cite{AlphaEdit} utilizes the null space for perturbation projections in knowledge editing, allowing new information to be integrated without overwriting or forgetting previously learned knowledge.
Consequently, in the context of language-guided recommendation systems, the null space represents a promising mechanism for incorporating collaborative signals while preserving the integrity of semantic information.

%% file: chapters/6-conclusion.tex
\section{Conclusion}

In this paper, we propose a language-guided learning strategy for ID embeddings, termed AlphaFuse, which effectively integrates collaborative signals while preserving semantic information. 
The core idea of AlphaFuse is to learn ID embeddings in the null space of language embeddings.
Specifically, we begin by performing singular value decomposition on language embeddings, which decomposes the semantic space into singular subspaces associated with distinct singular values. 
These subspaces are then categorized into two groups: the semantic-sparse null space and semantic-rich subspaces (\ie row space, complementary to the null space). 
Next, we perform targeted preprocessing to each subspace --- clipping the null space and standardizing the semantic-rich subspaces. 
Finally, we learn the ID embeddings within the post-clip null space, while preserving the semantic information in the standardized semantic-rich subspaces.
During inference, we obtain the final item representations by integrating the frozen language embeddings and the trained ID embeddings.
This approach preserves the original semantic space of language embeddings to prevent degradation, while ensuring the integration of language embeddings into the final item representations. 
Furthermore, it obviates the need for auxiliary trainable modules, seamlessly adapting to arbitrary sequential recommendation framework.
Through comprehensive experiments, we validate the effectiveness and flexibility of AlphaFuse.

\section*{Acknowledgments}
This research is supported by the National Natural Science Foundation of China (92270114, 12401375) and the MOE Project of Key Research Institute of Humanities and Social Sciences (22JJD110001). 
